\documentclass[preprint,times,10pt,3p,number]{elsarticle}
\usepackage{listings}
\title{Parallel FFT-based Poisson Solver for Isolated Three-dimensional Systems}
\author[utk]{Reuben D. Budiardja}
\ead{reubendb@utk.edu}

\author[utk,ornl]{Christian Y. Cardall}
\ead{cardallcy@ornl.gov}

\address[utk]{Department of Physics and Astronomy, University of Tennessee, Knoxville, TN 37996-1200, USA}
\address[ornl]{Physics Division and Computer Science and Mathematics Division, Oak Ridge National Laboratory, Oak Ridge, TN 37831-6354, USA}

\begin{document}

\begin{abstract}
We describe an implementation to solve Poisson's equation for an isolated system on a unigrid mesh using FFTs. 
The method solves the equation globally on mesh blocks distributed across multiple processes on a distributed-memory parallel computer. 
Test results to demonstrate the convergence and scaling properties of the implementation are presented. The solver is offered to interested users as the library \texttt{PSPFFT}.

\vspace{1pc}
\noindent
\textbf{Program summary}

\noindent
{\em Program Title:} PSPFFT \\
{\em Program obtainable from:} http://astro.phys.utk.edu/\_media/codes:pspfft-1.0.tar.gz, CPC Program Library\\
{\em CPC Catalogue identifier:} \texttt{AEJK\_v1\_0} \\
{\em Licensing provisions:} Standard CPC license, http://cpc.cs.qub.ac.uk/licence/licence.html \\
{\em Programming language:}  Fortran 95                                 \\
{\em Computer:} any architecture with a Fortran 95 compiler, distributed memory clusters  \\
{\em Operating system:} Linux, Unix\\
{\em RAM:} depends on the problem size, approximately $170$ MBytes for $48^3$ cells per process \\
{\em Has the code been vectorized or parallelized?:} Yes, using MPI \\
{\em Number of processors used:} arbitrary number (subject to some constraints). It has been tested from $1$ up to $\sim 13000$ processors. 
{\em Classification:} 4.3 Differential Equations   \\
{\em External routines/libraries:} 
  MPI (http://www.mcs.anl.gov/mpi/), FFTW (http://www.fftw.org), \\
Silo (https://wci.llnl.gov/codes/silo/) (only necessary for running test problem) \\
{\em Nature of problem:} Solving Poisson's equation globally on unigrid mesh distributed across multiple processes on distributed memory system. \\
{\em Solution method:} Numerical solution using multidimensional discrete Fourier Transform in a parallel Fortran 95 code. \\
{\em Unusual features:} This code can be compiled as library to be readily linked and used as a black-box Poisson solver with other codes. \\
{\em Running time:} Depends on the size of the problem, but typically less than 1 second per solve \\
\end{abstract}

\maketitle

\section{Introduction}
Some physics simulations require the solution of Poisson's equation with an isolated source distribution and a vanishing boundary condition at infinity. 
A common example is the calculation of the Newtonian gravitational potential of a self-gravitating astrophysical system. 
Poisson's equation is
\begin{equation}
\nabla^2 \Phi(\mathbf{x}) = S(\mathbf{x}), \label{eq:poisson}
\end{equation}
where $S(\mathbf{x})$ describes the known distribution of the source that generates the potential $\Phi ( \mathbf{x} )$. 
For instance, $S(\mathbf{x} )$ is proportional to the mass density in the context of Newtonian gravity, and to the charge density in electrostatics.

Several methods exist to solve the discretized Poisson's equation on a uniform grid. 
These include, for example, multigrid methods, iterative / relaxation methods, several matrix methods, and methods that employ Fourier transforms (for discussion of some of these, see for instance \cite{Hockney1989, Swarztrauber490, Dorr1970}). 
Here we implement, and extend to three dimensions, a particular method \citep{Hockney1970} (see also \cite{Hockney1989} and \cite{EASTWOOD1979}) of the latter class.
(Another well-known approach for isolated systems based on Fourier transforms \citep{James1977}, also discussed in \cite{Hockney1989}, would not be as straightforward to parallelize and is not discussed here.)
An advantage of this approach is that discrete Fourier transform algorithms have been well-studied, with the Fast Fourier transform (FFT) being the most commonly employed; it requires $O(n\log n)$ operations, where $n$ is the number of elements to transform.
Several FFT implementations, some freely available, also exist as libraries suitable for end-users. 

The key issue addressed by the implementation described here is the parallelization of an FFT-based algorithm for solving Poisson's equation for an isolated system. Obtaining such solutions in three dimensions requires resources that at present are most commonly available in distributed-memory parallel computers. 
Machines of this type allow large problems to be decomposed---for example, into multiple spatial subdomains---and distributed across different `processes' to be solved in parallel. 
(Each `process' contains its own copy of the program; can only access memory locations allocated either statically or dynamically by the program; and can communicate with other processes only through a specific protocol, the Message Passing Interface (MPI) \citep{Gropp1999} presently being the most widely used.) 
While Poisson's equation must be solved globally on the computational domain and across multiple processes, most FFT implementations require that all data be accessible on a single process; therefore data movement and coordination across multiple processes are key ingredients of our FFT-based approach. 

Our Poisson solver for isolated systems in three dimensions, named \texttt{PSPFFT} (`\texttt{P}oisson \texttt{S}olver \texttt{P}arallel \texttt{FFT}'), is available as a library that can readily be used by and linked to other programs, subject to a few constraints on numbers of processes and mesh points described in Subsec. \ref{subsubsec:decomposition}.
We use the FFTW library \citep{Frigo2005}, but our use of it is abstracted in such a way that other FFT libraries could be used without having to make widespread changes throughout the code. 
We use MPI to manage data movement and communication across processes, but our calls to message passing routines are abstracted as well. 

This paper is organized as follows. 
Section \ref{sec:SolutionMethod} outlines the algorithm as well as implementation details specific to our code. 
This is followed in Section \ref{sec:InstallationUsage} by instructions on installation and use of the program as a library.
Test problems illustrating the convergence properties and performance of our implementation are presented in Section \ref{sec:NumericalResults}.
Section \ref{sec:Conclusion} contains concluding remarks. 

\section{Solution Method} \label{sec:SolutionMethod}

\subsection{Formulation}

Our problem is to solve Eq. (\ref{eq:poisson}) with the boundary condition $\Phi(\mathbf{x}) \rightarrow 0$ as $|\mathbf{x}| \rightarrow \infty$ (vanishing boundary condition). Use of the Green's function
\begin{equation}
G(\mathbf{x}) = - \frac{1}{4\pi \left| \mathbf{x} \right| }, \label{eq:greens}
\end{equation}
which satisfies
\begin{equation}
\nabla^2 G(\mathbf{x}) = \delta (\mathbf{x}) 
\end{equation}
and obeys the vanishing boundary condition, yields the formal solution
\begin{equation}
\Phi(\mathbf{x}) = \int d\mathbf{x}'\; G(\mathbf{x}-\mathbf{x}')\,S(\mathbf{x}'). \label{eq:convolution}
\end{equation}
This integral may be evaluated with the help of the convolution theorem. Given the Fourier transforms $\tilde G(\mathbf{k})$ and $\tilde S(\mathbf{k})$ of $G(\mathbf{x})$ and $S(\mathbf{x})$, the Fourier transform of the potential is 
\begin{equation}
\tilde \Phi(\mathbf{k}) = \tilde G(\mathbf{k})\, \tilde S(\mathbf{k}).
\end{equation}
The potential $\Phi(\mathbf{x})$ is then obtained by an inverse Fourier transform.

When the Fourier transforms are to be done with FFTs, use of a regular mesh with the same spacings in each dimension is most natural; but in principle it should be possible to use any mesh for which a coordinate transformation can bring the mesh point positions to triplets of integers. For instance, to allow for a regular mesh with numbers of mesh points $n_x, n_y, n_z$ and unequal (but constant) mesh point spacings $h_x, h_y, h_z$ in the three dimensions, 
Eqs. (\ref{eq:greens})-(\ref{eq:convolution}) become
\begin{eqnarray}
G(\bar\mathbf{x}) &=& - \frac{h_x\,h_y\,h_z}{4\pi \sqrt{{h_x}^2 {\bar x}^2 + {h_y}^2 {\bar y}^2 + {h_z}^2 {\bar z}^2} }, \label{eq:greens2} \\
\left( \frac{1}{{h_x}^2}\frac{\partial^2}{\partial {\bar x}^2} + \frac{1}{{h_y}^2}\frac{\partial^2}{\partial {\bar y}^2} + \frac{1}{{h_z}^2}\frac{\partial^2}{\partial {\bar z}^2}\right) G(\bar\mathbf{x}) &=& \delta (\bar\mathbf{x}), \\
\Phi(\bar\mathbf{x}) &=& \int d\bar\mathbf{x}'\; G(\bar\mathbf{x}-\bar\mathbf{x}')\,S(\bar\mathbf{x}'), \end{eqnarray}
where the values of the transformed coordinates $\bar\mathbf{x}$ corresponding to the mesh points are triplets of integers ranging from $0$ to $n_x - 1,\ n_y - 1,\ n_z - 1$ in the three dimensions respectively. (Note that the Jacobian of the coordinate transformation has been absorbed into the numerator of Eq. (\ref{eq:greens2}), with the denominator still being $4\pi$ times the distance from the origin.)

\begin{figure}
  \centering
  \includegraphics[scale=0.3]{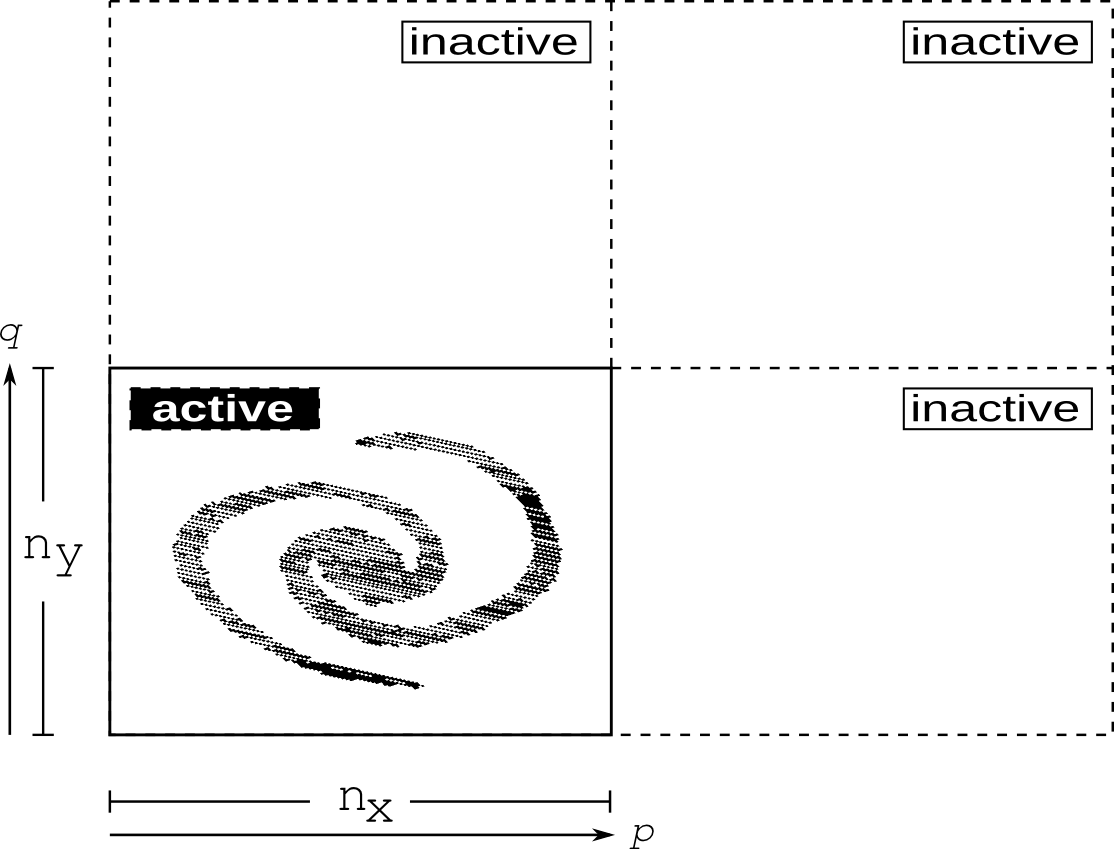}
  \caption{A two-dimensional illustration (redrawn after \cite{Hockney1970}) of the mesh doubling used to solve for the potential due to an isolated source. The `active' mesh on the lower left is the original computational domain with $n_x \times n_y$ cells.} 
  \label{fig:mesh_doubling}
\end{figure}

The implementation of boundary conditions at infinity on a necessarily finite computational domain 
can be handled `exactly', that is to say, with only discretization error, via a mesh doubling procedure and use of the standard periodic form of the discrete Fourier transform \citep{Hockney1970} (see also \cite{Hockney1989} and \cite{EASTWOOD1979}). 
Figure \ref{fig:mesh_doubling} illustrates this in two dimensions. 
The `active' portion of the mesh corresponds to the original computational domain, while the `inactive' portions are those arising from doubling the extent of the mesh in each dimension.
The source distribution is set to its known physical values in the active region, and to zero in the inactive regions.
Indexing cells by integer triplets $p,q,r$ (the position of the mesh points in transformed coordinates $\bar\mathbf{x}$), the Green's function in the active and inactive regions is 
\begin{eqnarray}
  \left. 
  \begin{array}{ll}
    G_{p,q,r} = - h_x\,h_y\,h_z \,(4\pi )^{-1}\left({h_x}^2 {p}^2 + {h_y}^2 {q}^2 + {h_z}^2 {r}^2\right)^{-1/2} \\
    G_{2n_x-p,q,r} = G_{p,2n_y-q,r} = G_{p,q,2n_z-r} = G_{2n_x-p,2n_y-q,2n_z-r} = G_{p,q,r} \\
    G_{2n_x-p,2n_y-q,r} = G_{2n_x-p,q,2n_z-r}  = G_{p,2n_y-q,2n_z-r}  = G_{p,q,r} \\
  \end{array} \nonumber 
   \right\} & 
  \begin{array}{rr}
    0 \le p, q, r \le n_x, n_y, n_z\\
    p+q+r \ne 0
  \end{array} \nonumber \\
  G_{0,0,0} = - h_x\,h_y\,h_z \, \left[4\pi \; \mathrm{min}(h_x,h_y,h_z) \right]^{-1}.  &&
\label{eq:greens_function}
\end{eqnarray}
That is, the Green's function in the active region follows Eq. (\ref{eq:greens2}), and is set up in the inactive regions in such a way that a periodic replication of the doubled mesh in all dimensions yields Eq. (\ref{eq:greens2}) in the entire region $-n_x,-n_y,-n_z \le p,q,r \le n_x, n_y, n_z$ surrounding the origin. The value for $G_{0,0,0}$ regularizes the singularity at the origin by assigning it to be equal to the largest off-origin value on the mesh; for $h_x = h_y = h_z = 1$, this reduces (up to conventions for sign and the $4\pi$ factor) to the prescription given in \cite{Hockney1970}.

\subsection{Program Implementation}

\texttt{PSPFFT} is written in Fortran 95, using an object-oriented programming style to a degree convenient and possible within that language. The library FFTW \citep{Frigo2005} provides our basic FFT functionality, and we use MPI \citep{Gropp1999} to implement parallelization across multiple processes. 

\subsubsection{Domain Decomposition and Storage} \label{subsubsec:decomposition}
\begin{figure}
  \centering
  \includegraphics[scale=0.3]{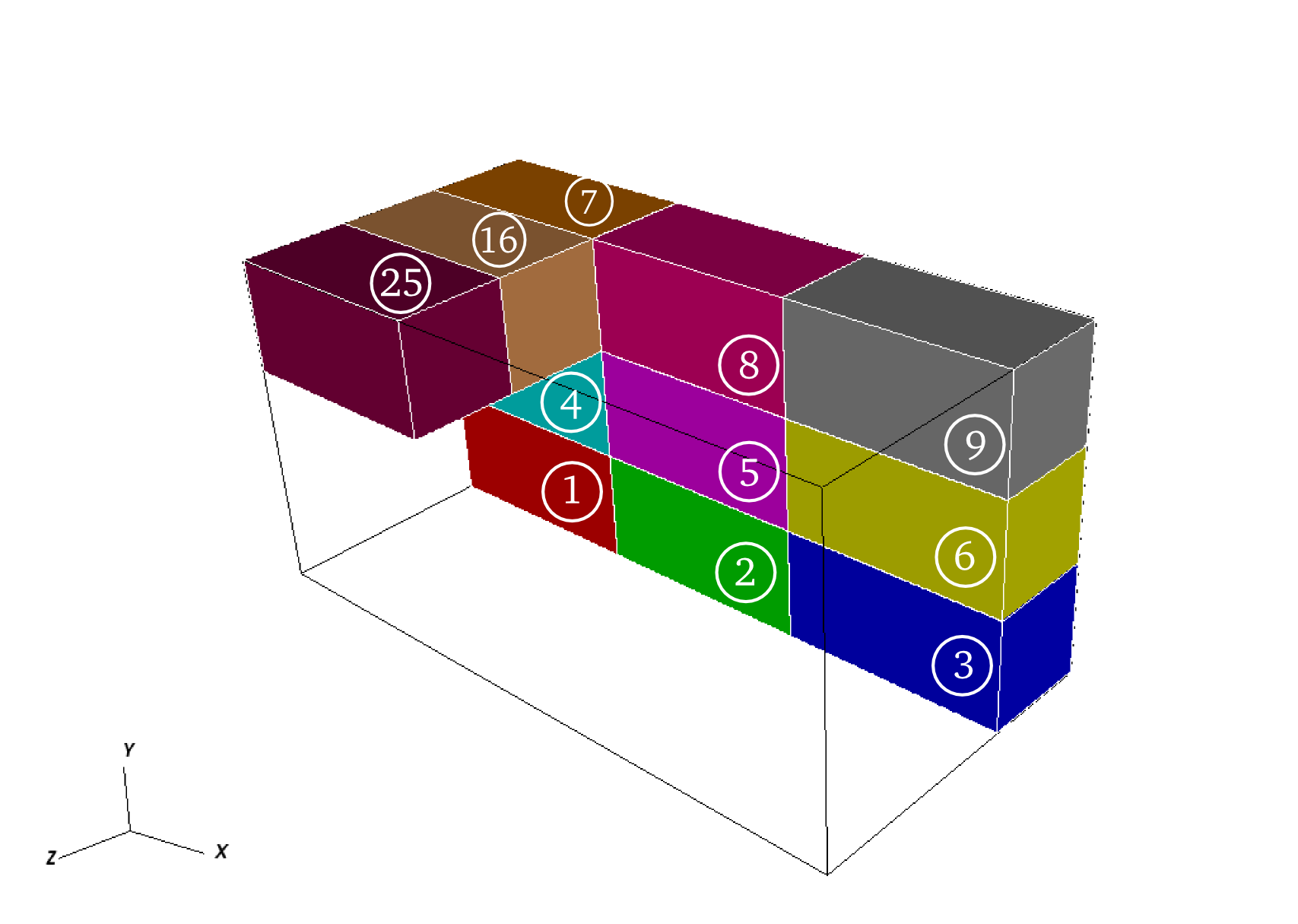}
  \caption{An illustrative brick decomposition in three dimensions for a computational domain assigned to twenty-seven processes. Only eleven bricks are shown to simplify the illustration. The bricks are labeled with the rank of the process that `owns' them. (Process rank numbering here and in the following two figures begins with 1, rather than 0 as in MPI and internally in the code.)} 
  \label{fig:domain_decomposition}
\end{figure}

In many physical simulations the problem size is large enough that the computational domain is spatially decomposed into multiple subdomains, each assigned to a different process. 
In the general case the extent of the domain---and/or the number of mesh points---need not be the same in all dimensions. 
For the purpose of minimizing communications, decompositions yielding subdomains with low surface-to-volume ratio are favorable in situations (such as hydrodynamics) in which nearest-neighbor information is required. 
Our code assumes that the source $S(\mathbf{x})$ of Eq. (\ref{eq:poisson}) is available, and that the potential $\Phi(\mathbf{x})$ is desired, in a simple `brick' decomposition: in three dimensions, the computational domain is divided in each dimension by $n_b=\sqrt[3]{n_p}$, the cube root of the number of processes $n_p$. 
For simplicity we require $n_p = {n_b}^3$ to be a perfect cube (in three dimensions), and the number of mesh points in each dimension to be evenly divisible by $n_b$. 
Figure \ref{fig:domain_decomposition} illustrates the brick decomposition.

\begin{figure}
  \centering
  \includegraphics[scale=0.4]{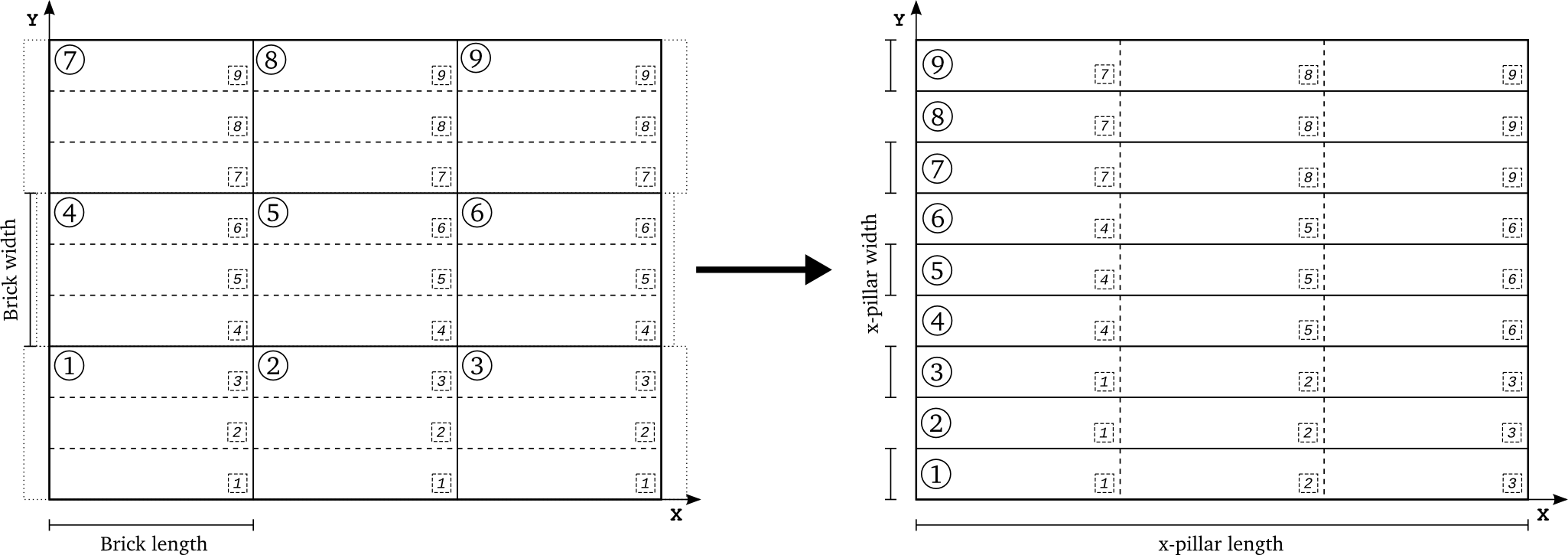}
  \caption{An illustration of the transformation from brick decomposition to $x$ pillars from a three-dimensional mesh assigned to twenty-seven processes. 
Here only the first (lowest in the $z$ direction) $xy$ slab of bricks is shown; other slabs independently follow the same transformation. 
The left panel shows the whole computational domain decomposed into bricks, demarcated by solid lines and assigned to processes labeled by the numbers in solid circles. 
Dashed lines in the left panel mark the chunks of data that need to be sent to the processes labeled by the numbers in dashed squares in order to build the pillars. 
As indicated by the dotted boundaries, processes $[1,2,3]$, $[4,5,6]$, and $[7,8,9]$ form three separate groups, each with its own subcommunicator within which chunks of data are exchanged during the construction of the $x$  pillars. 
In the right panel, we see that each process (again, labeled by numbers in solid circles) also owns a pillar. 
The boundaries between pillars are now marked by solid lines, and the dashed lines indicate the chunks of data that the process received from other processes labeled by numbers in dashed squares.}  
  \label{fig:bricks_to_pillars}
\end{figure}

The brick decomposition is not convenient for FFTs, however, because a single transform is most naturally and efficiently performed on data accessible to a single process; therefore our solver has its working storage arranged in `pillars' rather than bricks. 
(A distributed parallel FFT that leaves data in place in the brick decomposition---and therefore requires parallelized one-dimensional FFTs---has also been mentioned in the literature \cite{BUSH2006}. In contrast, our approach allows in principle the use of any of several widely available and highly optimized serial FFT libraries.)
The `length' of what we term `$x$ pillars' spans the full extent of the computational domain in the $x$ direction. 
The `width' of the $x$ pillars is their extent in the $y$ direction, which is $1/n_b$ times the $y$ extent of the bricks. This implies another constraint imposed by our solver: the number of mesh points $n_y / n_b$ spanned by the $y$ extent of a brick must itself be evenly divisible by $n_b$.    
The `height' of the $x$  pillars, which is their extent in the $z$ direction, is the same as the extent of the bricks in the $z$ direction. 
By similar construction (and with similar constraints on $n_z$ and $n_x$), we have $y$ pillars and and $z$ pillars whose (width, height) are taken to be their extents in $(z, x)$ and $(x, y)$ respectively. 
These `pillar decompositions' cover the same total volume and contain the same total number of mesh points as the brick decomposition, as illustrated in Fig. \ref{fig:bricks_to_pillars}.
Finally, we note that the lengths of the pillars are doubled as necessary to accommodate the mesh doubling procedure, so that the pillars span both the active and inactive portions of the mesh.

Because of the row-major nature of Fortran array storage, a pillar's length, width, and height correspond in our code to the first, second, and third dimensions of a rank-three array. 
This allows a $(\mbox{width} \times \mbox{height})$-number of one-dimensional FFTs to be performed efficiently on contiguous data, specifically on lines containing a number of data points equal to the pillar length. 
The construction of pillars from bricks and vice-versa requires data movement across different processes.
Using MPI, this is accomplished by creating a subcommunicator for each group of processes that need to communicate data among themselves, as illustrated in Fig. \ref{fig:bricks_to_pillars}. 
For each group, a call to \texttt{MPI\_AllToAll} can then be made with the group's subcommunicator in order  to achieve the construction of pillars. 
These subcommunicators are saved to be used in the reverse process of deconstructing pillars back into bricks.

\subsubsection{Multidimensional Transforms}
A multidimensional FFT can be accomplished as a sequence of sets of one-dimensional transforms. 
The number of required operations is still of $O(n\log n)$, where $n = n_x n_y n_z$ is the total number of mesh points.
One possibility for achieving computational efficiency is to transpose data between transforms in subsequent dimensions in order to achieve contiguity of memory access in each dimension. 
In any case, such transpose operations become a necessity in a distributed memory environment if parallelization of individual one-dimensional transforms is to be avoided.

\begin{figure}
  \centering
  \includegraphics[scale=0.4]{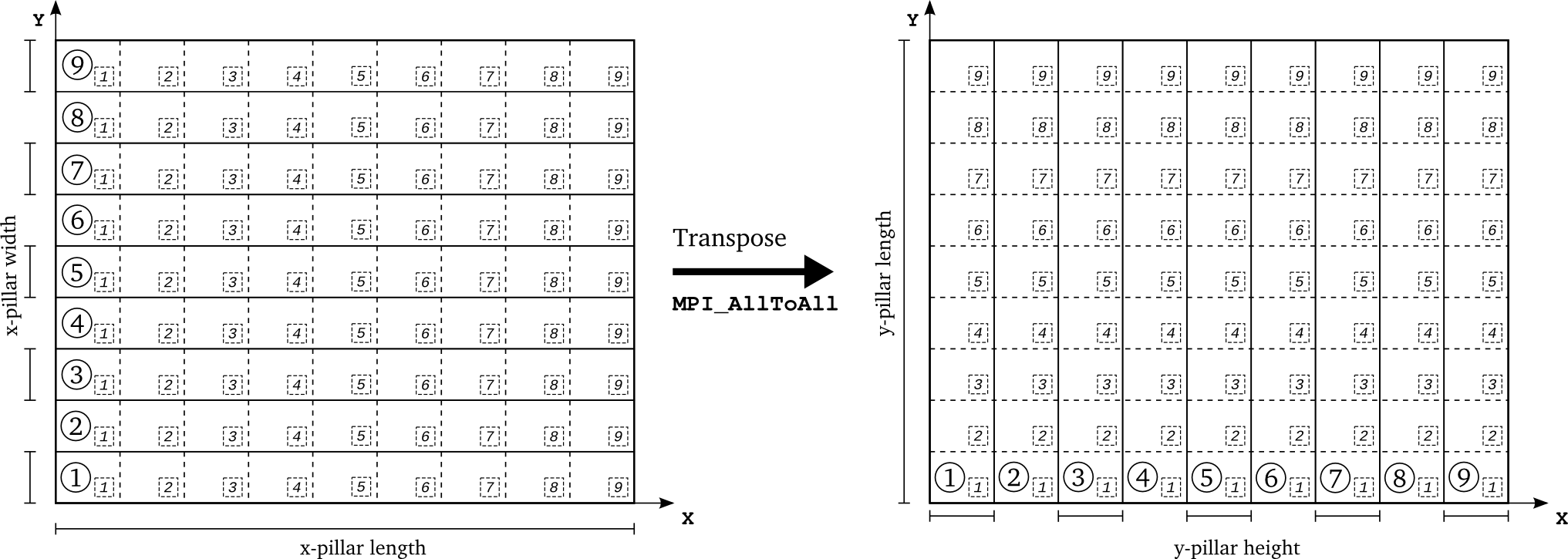}
  \caption{An illustration of the transpose of $x$ pillars to $y$ pillars on a three-dimensional mesh assigned to twenty-seven processes. As before, only the first (lowest in the $z$ direction) $xy$ slab is shown. The solid rectangles demarcate data owned by different processes, labeled by numbers in solid circles. Dashed lines mark chunks of data that need to be sent to (left panel) and received from (right panel) processes labeled by numbers in dashed boxes. In this example, a slab with 9 processes forms a single MPI group with its own subcommunicator, and the transpose is accomplished with a call to \texttt{MPI\_AlltoAll}.}
  \label{fig:x-pillars_to_y-pillars}
\end{figure}

The sequence of transforms and transposes is as follows.
Data are initially loaded into the solver's $x$ pillars: during initialization the Green's function is set up directly in the $x$ pillars according to Eq. (\ref{eq:greens_function}), while the source is transferred from the brick decomposition to the $x$ pillars at the beginning of every solve.
With data loaded in $x$ pillars, multiple one-dimensional transforms in the $x$ dimension are simultaneously performed by all processes. 
The $y$ pillars are then populated, independently in separate $xy$ `slabs', as illustrated in Fig.  \ref{fig:x-pillars_to_y-pillars}.
For each slab a separate MPI group with its own subcommunicator is created;  
thus are are $n_b=\sqrt[3]{n_p}$ subcommunicators, each of which has ${n_b}^2$ processes.
Within each subcommunicator a call to \texttt{MPI\_AllToAll} transposes the data from $x$  pillars to $y$  pillars so that FFTs can be performed in the $y$  direction. 
Similar transposes in $yz$ slabs allow FFTs to be performed in $z$ pillars. Here the multiplication of the transforms of the Green's function and the source takes place as well,
with the resulting Fourier-space solution of the Poisson equation overwriting the transformed source distribution. 
A reverse sequence of backward transforms and transposes gets the solution (modulo a normalization factor due to the multiple transforms) back into real space, stored in the $x$ pillars. 
Finally the solution is redistributed from the active portion of the $x$ pillars to the brick decomposition, overwriting the storage in which the source was delivered.

This sequence of transforms and transposes makes use of permanent storage for the source distribution in $x$, $y$, and $z$ pillars, which at the end of the solve is reused to store the potential. 
This same storage is then updated with a new source distribution on the next call to the Poisson solver. 
The transform of the Green's function is computed only once, and stored permanently in $z$ pillars, when the solver is initialized. 
Computation of the transformed Green's function requires $x$  pillars and $y$  pillars, but these are deallocated at the end of the initialization.

\section{Installation and Usage} \label{sec:InstallationUsage}

\subsection{Installation}
\texttt{PSPFFT} is distributed as a gzip-compressed \texttt{.tar} file. 
Upon uncompression and extraction, the top-level directory \texttt{PSPFFT} has a \texttt{README} file and four subdirectories: \texttt{Build}, \texttt{Config}, \texttt{Source}, and \texttt{Test}. 
Complete installation instructions are given in the \texttt{README} file. 
During the build process, object files (with a \texttt{.o} extension) and Fortran module files (usually with a \texttt{*.mod} extension) are placed in the \texttt{Build} directory.  
The \texttt{Config} directory includes a configuration file in which system-specific values for the build and installation process are specified; 
these include the name of the MPI compiler wrapper, the location of the FFTW library, and the location of the installation.
The directories \texttt{Source} and \texttt{Test} contain the \texttt{PSPFFT} source code and a test program respectively. 
The test program solves the problem described in Sec. \ref{subsec:homogeneous_spheroid}. 

Brief installation instructions are as follows. 
After replacing the system-specific values in the file \texttt{Config/Makefile\_Config}, the user should change to the \texttt{Build} directory to build and install the library by typing
\begin{verbatim}
> make
> make install 
\end{verbatim} 
This creates and copies the library file \texttt{libpspfft.a} and the Fortran module files to the installation directory. 
The library file can then be linked to programs that need the solver. 

\subsection{Usage}
This code snippet and the explanation that follows illustrate how to use the \texttt{PSPFFT} library in a Fortran program. 
\begin{lstlisting}[language=fortran,basicstyle=\footnotesize, numbers=left, numberstyle=\footnotesize,stepnumber=1,numbersep=7pt,frame=tb,xleftmargin=10pt]
program Main
  use PSPFFT 
  implicit none
  include 'mpif.h'
  
  integer :: &
    Error
  integer, dimension(3) :: &
    nCellsPerProcess, &
    nTotalCells 
  real(KR), dimension(3) :: &
    CellWidth 
  type(ArrayReal_3D_Base), dimension(1) :: &
    SourceSolution
  type(PSPFFT_Form), pointer :: &
    PoissonSolver

  call MPI_INIT(Error)
  !-- Add lines to set nCellsPerProcess, nTotalCells, and CellWidth
  call Initialize(SourceSolution, nCellsPerProcess)
  !-- Add lines to fill in SourceSolution with the source distribution
  call Create(PoissonSolver, CellWidth,  nTotalCells, MPI_COMM_WORLD, &
              VerbosityOption = CONSOLE_INFO_2)
  call Solve(PS, SourceSolution)
  !-- Add lines that use the potential returned in SourceSolution
  call Destroy(PS)
  call Finalize(SourceSolution)
  call MPI_FINALIZE(Error)
end program Main
\end{lstlisting}
All the derived data types, parameters, and subroutines used in the above example (except MPI-related variables and subroutines, such as \texttt{MPI\_COMM\_WORLD} and \texttt{MPI\_INIT}) are defined by the Fortran module \texttt{PSPFFT}, which is used by the main program on line $2$. 
After initializing MPI, the user should specify the number of cells per process,  the total number of cells across all processes, and the cell width---all of which are arrays specifying these quantities in each of the three dimensions---as indicated on line $19$. 

The variable \texttt{SourceSolution} is an array of derived data type \texttt{ArrayReal\_3D\_Base}. 
This is essentially a facility to make an array of rank-three arrays. 
\texttt{ArrayReal\_3D\_Base} has the following definition:
\begin{lstlisting}[language=fortran,basicstyle=\footnotesize, xleftmargin=10pt]
type, public :: ArrayReal_3D_Base
  real(KR), dimension(:,:,:), allocatable :: &
    Data
end type ArrayReal_3D_Base
\end{lstlisting}
The first, second, and third dimensions in \texttt{SourceSolution\%Data} correspond to the $x$, $y$, and $z$  dimensions on the mesh subdomain owned by a particular process.
Its value specifies the source in that particular cell.
Therefore this variable initially specifies the source distribution on the mesh which should be filled in by user.
Line $20$ simply initializes \texttt{SourceSolution} by allocating the necessary storage.

The call to the \texttt{Create()} subroutine on line $22$ allocates storage for the \texttt{PoissonSolver} variable and does all the necessary initializations. 
An optional argument \texttt{VerbosityOption} controls the verbosity of the messages the solver prints to \texttt{STDOUT}.
Acceptable values (defined as parameters), in decreasing order of verbosity, are \texttt{CONSOLE\_INFO\_2}, \texttt{CONSOLE\_INFO\_1}, \texttt{CONSOLE\_WARNING}, and \texttt{CONSOLE\_ERROR}.
If the argument is omitted, the verbosity level is set to \texttt{CONSOLE\_ERROR} (the least verbose) by default.
It is possible to replace the fourth argument (i.e. \texttt{MPI\_COMM\_WORLD} in this example) with a different MPI communicator which specifies a subgroup of processes that should be involved in solving the Poisson equation, subject to the constraints on numbers of processes and mesh points described in Section  \ref{subsubsec:decomposition}.
 
Line $24$ solves the Poisson equation with the source distribution passed as an argument. 
Upon return, the variable for the source is overwritten with the values of the potential.

The allocatable storage is deallocated by the calls to \texttt{Destroy()} and \texttt{Finalize} in lines $26-27$. 

All public subroutines exposed by \texttt{PSPFFT} are defined as generic interfaces using the function and subroutine overloading feature of Fortran 95. 
Therefore, other subroutines with the same names may be defined and will not conflict with the library as long as they are defined as generic interfaces. 

Assuming the above code example is in the file \texttt{Main.f90}, compilation and linking to \texttt{PSPFFT} are accomplished by a command like the following:
\begin{verbatim}
> mpif90 -L $(INSTALL)/lib -l pspfft -I $(INSTALL)/include Main.f90
\end{verbatim}
where \texttt{\$(INSTALL)} is the installation location of the library, and 
\texttt{mpif90} may also need to be replaced by a system-specific MPI compiler wrapper.

\section{Numerical Results} \label{sec:NumericalResults}
In this section we present a few illustrative test problems, investigate the numerical convergence of our code with respect to mesh resolution, and explore its scaling behavior on a distributed-memory parallel computer. 
The chosen test problems are broadly similar to systems encountered in astrophysical simulations, except that they have analytic solutions; this allows us to verify the correctness of our implementation. 

\subsection{Gravitational potential of spherical uniform mass}
\label{subsec:spherical_uniform_mass}
Consider the gravitational potential of a sphere of radius $R$ and uniform mass density $\rho$. 
The source is $4\pi G \rho$, and the potential as a function of radius $r$ from the center of the sphere has a simple analytic solution:
\begin{equation}
\Phi(r) = \left\{ 
\begin{array}{ll}
 - \displaystyle\frac{2}{3} \pi G \rho \left(3 R^2 - r^2 \right) 
     &\mbox{for $r \leq R$}\\                
 -  \displaystyle\frac{4}{3} \pi G \rho \displaystyle\frac{R^3}{r}
     &\mbox{for $r > R$}.\\
\end{array} \right. 
\end{equation}

\begin{figure}
  \centering
  \includegraphics[scale=0.2]{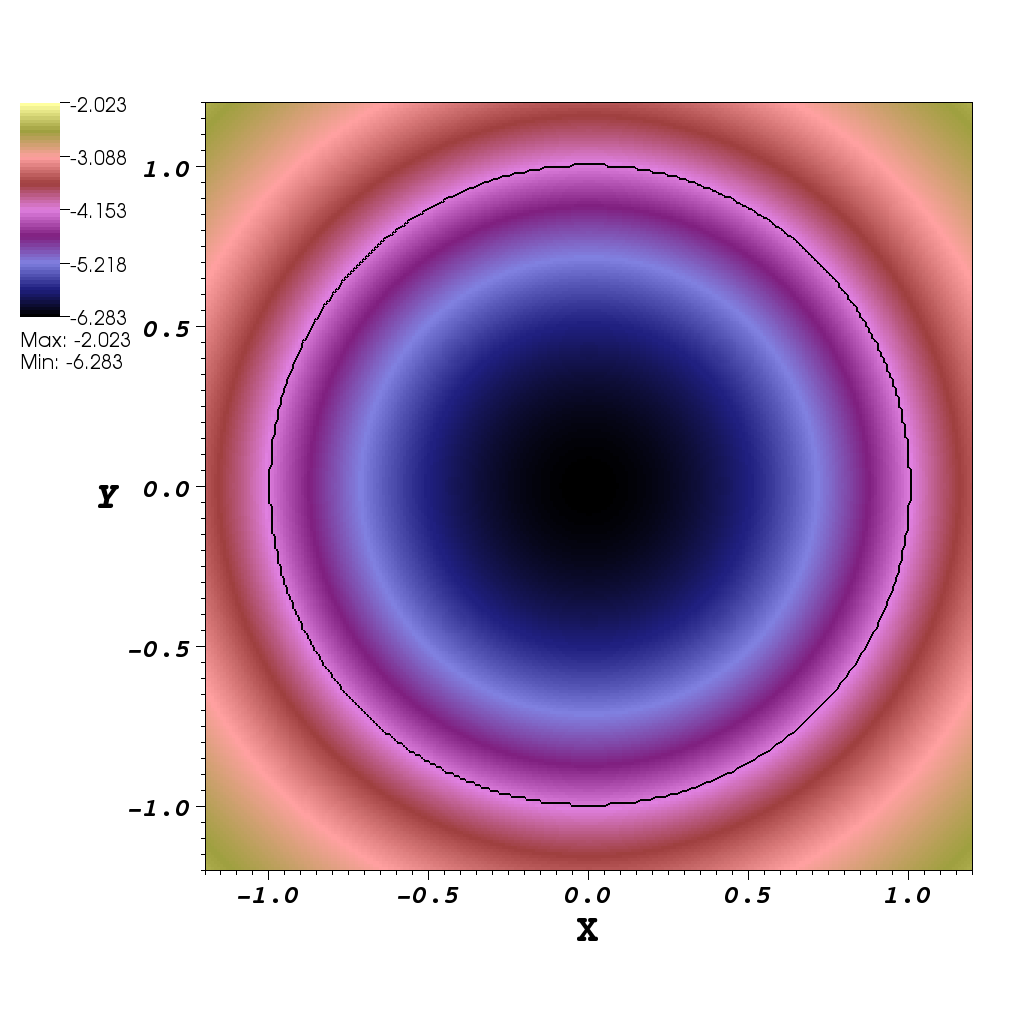}
  \caption{The potential of a unit sphere with uniform mass density $\rho=1$. 
  The figure is a slice through the three-dimensional mesh crossing the origin to show the $XY$-plane. 
  The solid black line indicates the surface of the sphere at radius $R=1$. The mesh resolution is 256 cells in each dimension.
  \label{fig:uniform_mass_potential}}
\end{figure}

We compute the potential for a sphere of radius $R=1$ and mass density $\rho=1$ in a Cartesian computational box with inner and outer boundaries at $[-1.2,+1.2]$ respectively in all dimensions. 
The sphere is centered on the origin of the coordinate system. 
For each mesh resolution, we calculate the potential in two ways: first, by using the analytic solution above with $r = \sqrt{x^2 + y^2 + z^2}$, and second, by using our implementation of the Poisson solver. 
By varying the mesh resolution we can check the convergence properties of our solver with respect to spatial resolution. 
The potential for this test problem is shown in Figure \ref{fig:uniform_mass_potential}.

\begin{figure}
  \centering
  \includegraphics[scale=0.2]{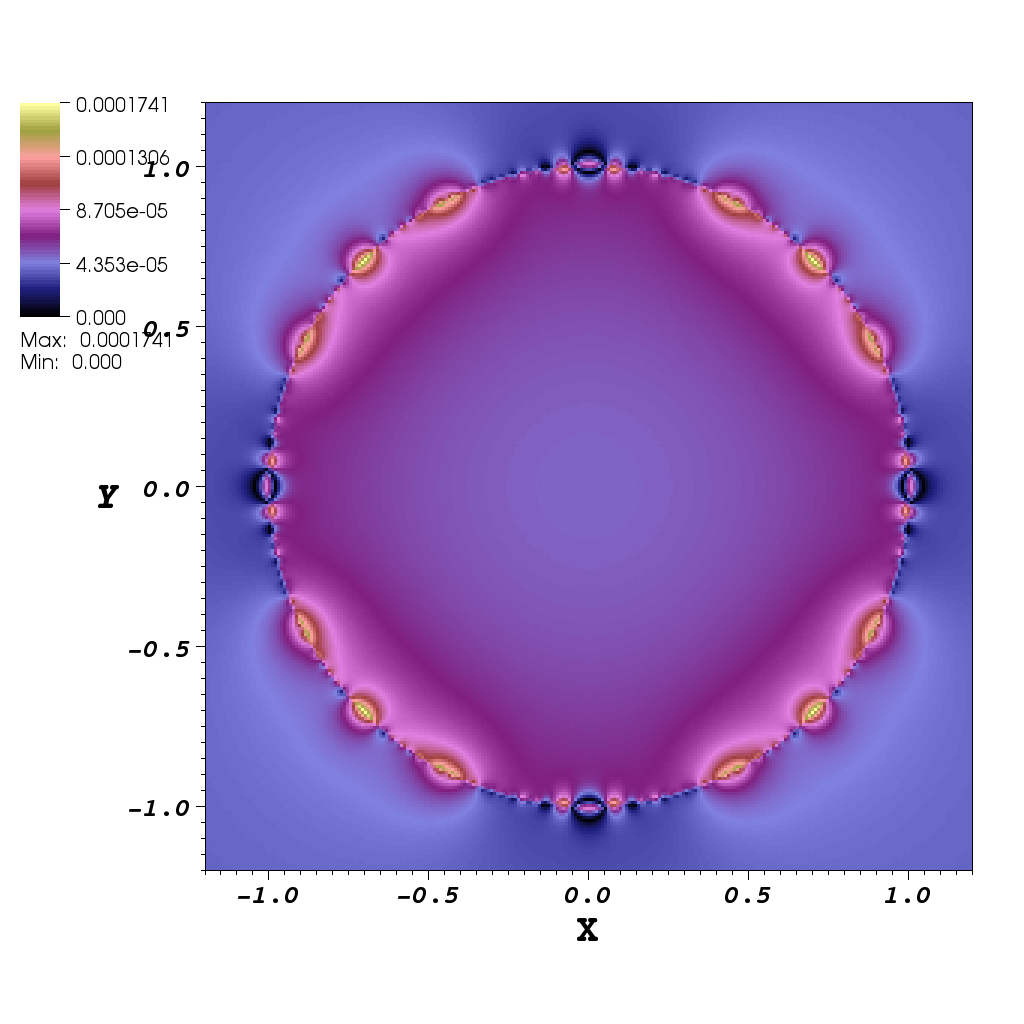}
  \caption{The relative error, as described by equation \ref{eq:L1-Norm_Error} but without the summation over all cells, of the potential of a spherical uniform-density mass. 
The figure is a slice through the mesh showing the $XY$-plane. 
The mesh resolution is 256 cells in each dimension. 
The largest errors are on the surface of the sphere due to the nature of the Cartesian grid. \label{fig:uniform_mass_error}}
\end{figure}

We use the usual definition of the $L_1$ norm to measure the error of the potential $\Phi$ computed by our solver relative to the analytic solution $\Phi_0$:
\begin{equation}
 L_{1} \left( \Phi \right) 
   = \frac{\displaystyle{\sum_{i,j,k}} \left| \Phi \left( x_i,y_j,z_k \right)  
                               - \Phi_0 \left( x_i,y_j,z_k \right) \right| } 
          {\displaystyle{\sum_{i,j,k}} \left| \Phi_0 \left( x_i,y_j,z_k \right) \right| }.
  \label{eq:L1-Norm_Error}
\end{equation} 
The summation is over all cells, indexed by $i$, $j$, $k$,. The $L_1$ norm gives a single number as a quantitative measure of the error for a given mesh resolution; in contrast, 
Figure \ref{fig:uniform_mass_error} illustrates the distribution of the relative error on the grid for a particular resolution, which is representative (by different constant factor) of the error distribution for other resolutions. 

\begin{figure}
  \centering
  \includegraphics[scale=0.5]{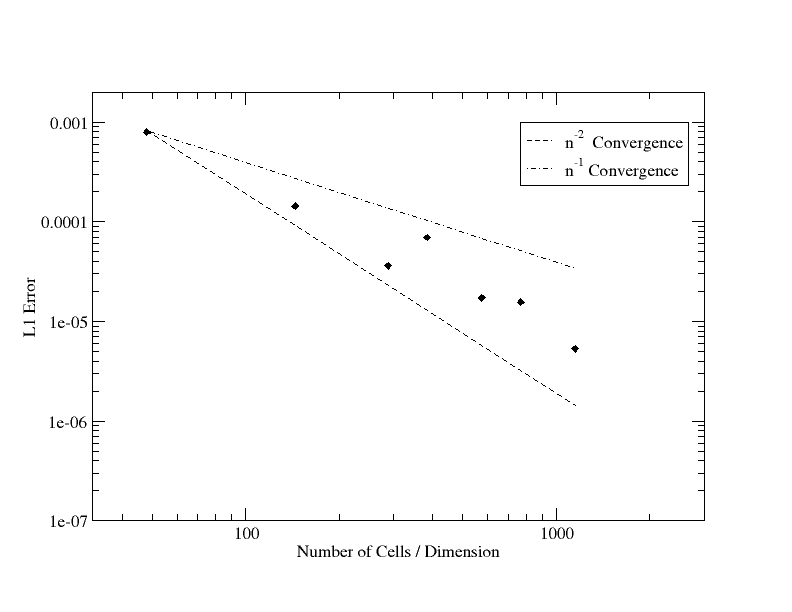}
  \caption{Convergence test for the potential of a spherical uniform-density mass. 
The $L_1$-norm relative error of the computed potential as compared to the analytical solution is plotted as function of the following mesh resolutions: $[48^3,144^3,288^3,384^3,576^3,768^3,1152^3]$. 
The dot-dashed and dashed lines are references for first- and second-order error convergence respectively.}
  \label{fig:sphere_convergence}
\end{figure}

Figure \ref{fig:sphere_convergence} shows the convergence of our solver (relative error as a function of resolution) for this problem. 
The convergence of the error trend is better than first order. 

\subsection{Gravitational potential of homogeneous spheroid} \label{subsec:homogeneous_spheroid}
A more general case of the previous test problem is the potential of a spheroid with uniform density $\rho$. 
The spheroid is formed by an ellipse centered at the origin and rotated about the $z$  axis, and is described by the equation
\begin{equation}
\frac{x^2+y^2}{a^2} + \frac{z^2}{b^2} = 1, 
\end{equation}
where $a$ and $b$ are the semi-diameters of the spheroid. 
The spheroid is oblate when $a > b$, with eccentricity $e$ defined as 
\begin{equation}
e = \sqrt{1-\left( \displaystyle\frac{b}{a}\right)^2}.
\end{equation}
The gravitational potential of a homogeneous spheroid \citep{Ricker2008} is a simpler case of the potential of a homogeneous ellipsoid given in \cite{Chandrasekhar1987}. 
Inside the spheroid, 
\begin{equation}
\Phi(x,y,z) = -\pi G \rho \left[ A \left(2a^2-x^2 -y^2\right) + B \left(b^2 - z^2\right) \right],
\label{eq:spheroid_potential_inside}
\end{equation}
where
\begin{equation}
  A = \displaystyle\frac{\sqrt{1-e^2}}{e^3} \, \mbox{sin}^{-1} e  - \frac{1-e^2}{e^2},
\end{equation}
\begin{equation}
  B = \displaystyle\frac{2}{e^2}-\displaystyle\frac{2\sqrt{1-e^2}}{e^3}\, \mbox{sin}^{-1} e.
\end{equation}
Outside the spheroid the potential is given by 
\begin{equation}
\Phi(x,y,z) = -2 \pi \rho G \displaystyle \frac{ab}{e}
                \left[\mbox{tan}^{-1} h -\frac{1}{2 a^2 e^2} \left\{ \left(x^2+y^2\right)
                                         \left( \, \mbox{tan}^{-1} h  - \frac{h}{1+h^2} \right)
                       + 2z^2\left(h-\,\mbox{tan}^{-1} h \right) \right\} \right],
\label{eq:spheroid_potential_outside}
\end{equation} 
with
\begin{equation}
h \equiv \frac{a \, e}{\sqrt{b^2+\lambda}},
\end{equation}
in which $\lambda$ is the positive root of the equation
\begin{equation}
\frac{x^2+y^2}{a^2+\lambda} + \frac{z^2}{b^2+\lambda} = 1.
\label{eq:spheroid_potential_auxiliary}
\end{equation}

\begin{figure}
  \centering
  \includegraphics[scale=0.2]{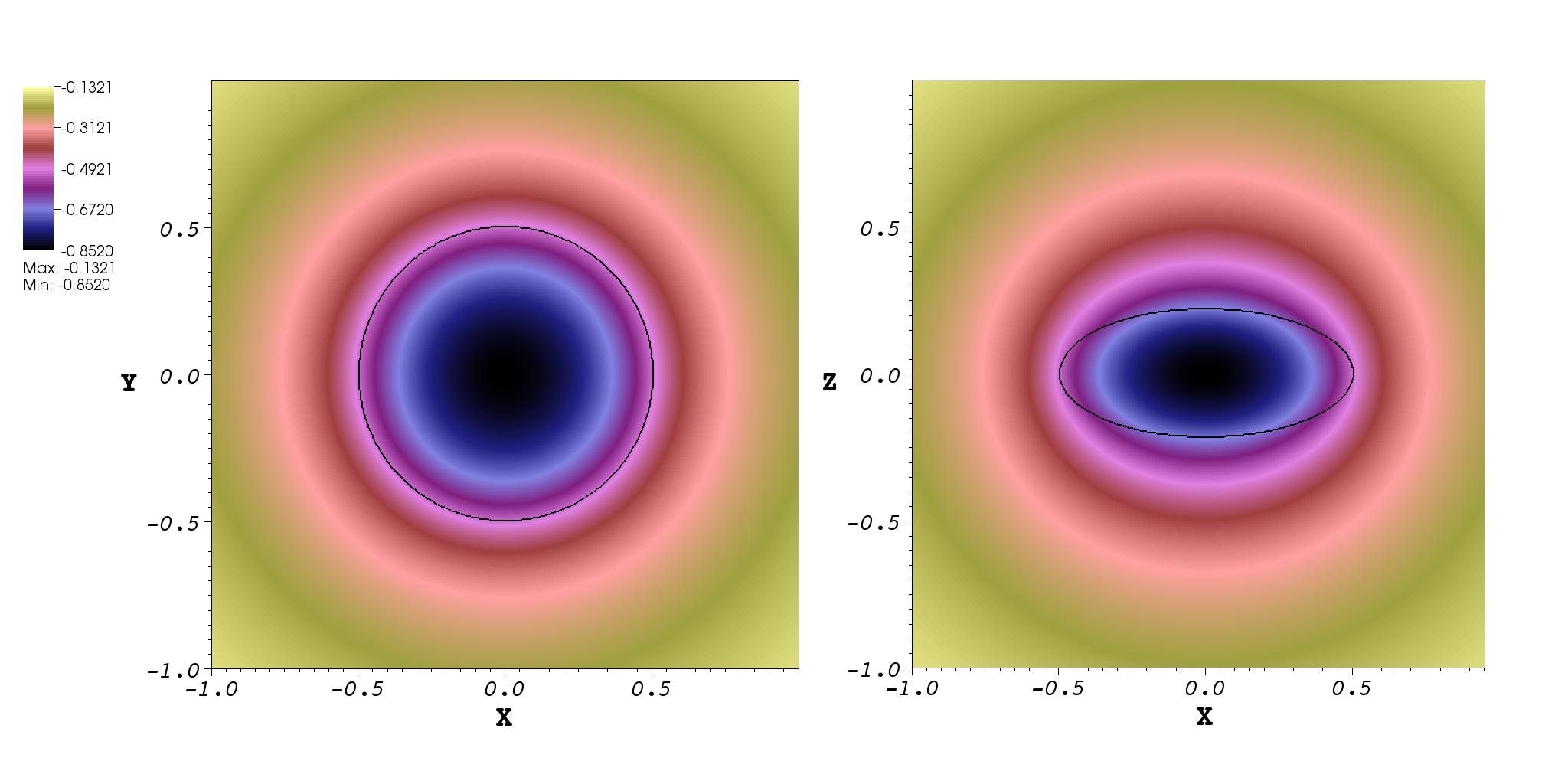}
  \caption{The potential of homogeneous spheroid with eccentricity $e=0.9$ and semi-major axis $a=0.5$ on a mesh with spatial resolution of $384$ cells in each dimension. 
The figures are slices of the mesh through the origin showing both the $XY$- and $XZ$-plane. 
The solid black line indicates the surface of the spheroid. 
  \label{fig:spheroid_potential}}
\end{figure}

We compute the potential for a spheroid with eccentricity $e=0.9$ and semi-major axis $a=0.5$ on a Cartesian computational box of size two in each dimension. 
As before, we set $\rho=1$. Figure \ref{fig:spheroid_potential} shows the computed potential for a particular mesh resolution. 

\begin{figure}
  \centering
  \includegraphics[scale=0.2]{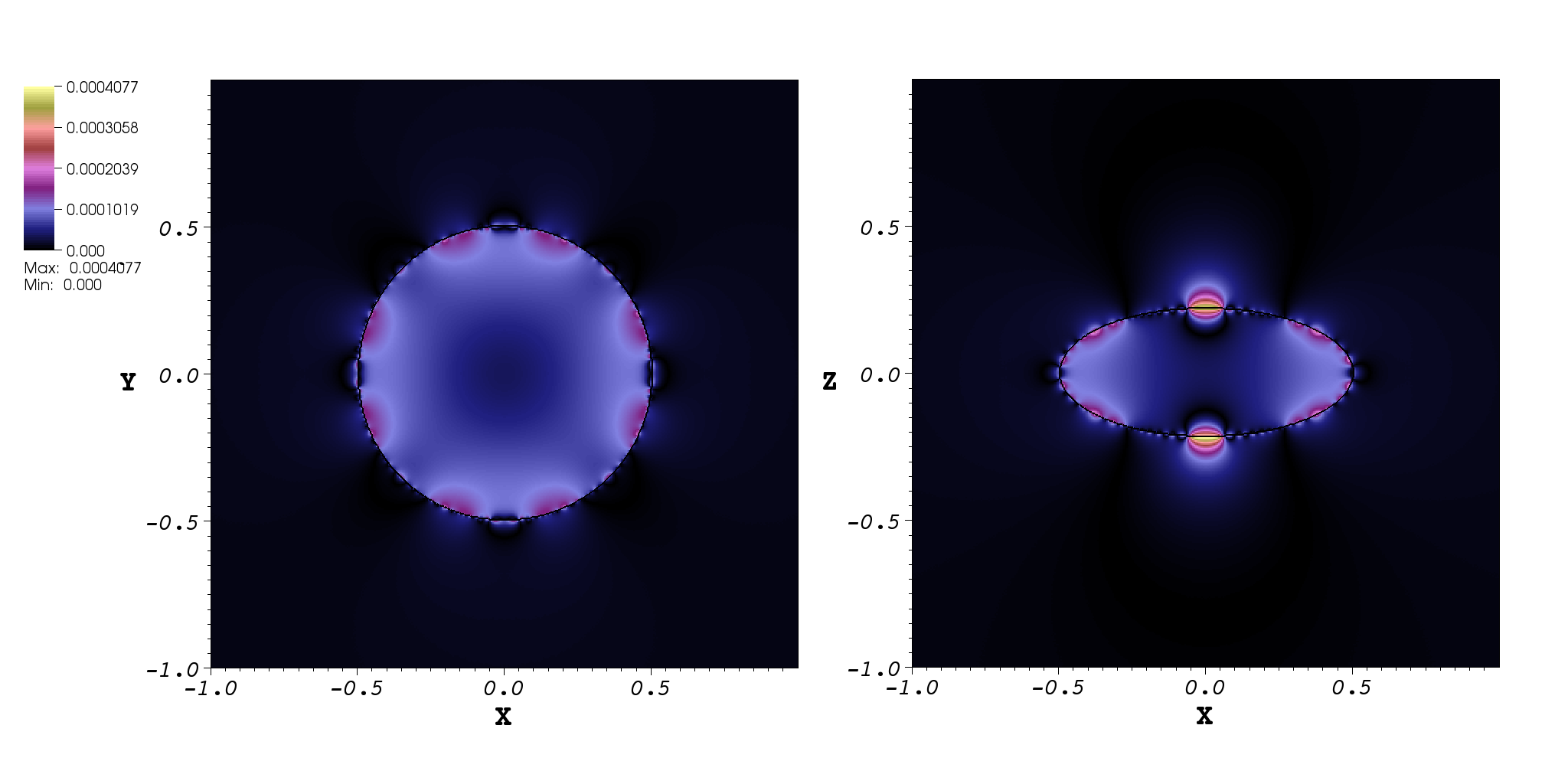}
  \caption{The relative error distribution of the homogeneous spheroid on a mesh with resolution of $384$ cells in each dimension. 
  Slices showing the $XY$- and $XZ$-planes are shown. 
  As before, the solid black line indicates the surface of the spheroid. 
  \label{fig:spheroid_error}}
\end{figure}

\begin{figure}[!ht]
  \centering
  \includegraphics[scale=0.4]{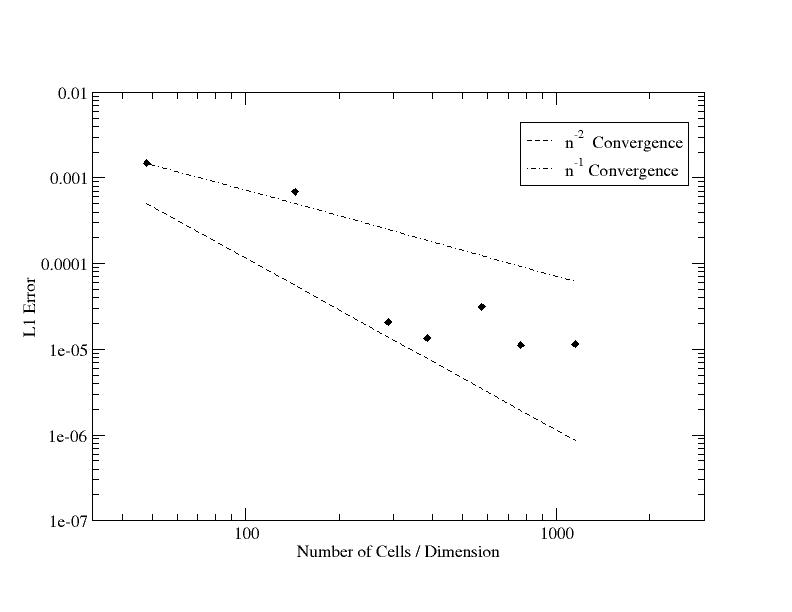}
  \caption{Convergence test of potential for a homogeneous spheroid with mesh resolutions $[48^3,144^3,288^3,384^3,576^3,768^3,1152^3]$.}
  \label{fig:spheroid_convergence}
\end{figure}

As in the previous test problem, we consider the error of the numerical solution relative to the analytic solution.
Figure \ref{fig:spheroid_error} illustrates the spatial distribution of the error for a particular resolution. 
The convergence of the error (specifically, the $L_1$-norm) with higher resolution is shown in Fig. \ref{fig:spheroid_convergence}; we see that on this problem our solver has higher than first order convergence, but less than second order. 

\subsection{Gravitational potential of homogeneous binary spheroid}
\label{subsec:binary_spheroid}
In this problem we place two separate homogeneous spheroids in the computational domain. 
The extent of the domain in the $x$ direction is twice that of the previous test problem, so that the $x$ dimension has inner and outer boundaries at coordinates $\pm 2$. 
The spheroids are centered on coordinates $\left[\pm1,0,0\right]$. In order to maintain the same effective resolution as our previous test problem, we double the number of cells in $x$ dimension only, resulting in a rectangular computational box. 

\begin{figure}
  \centering
  \includegraphics[scale=0.3]{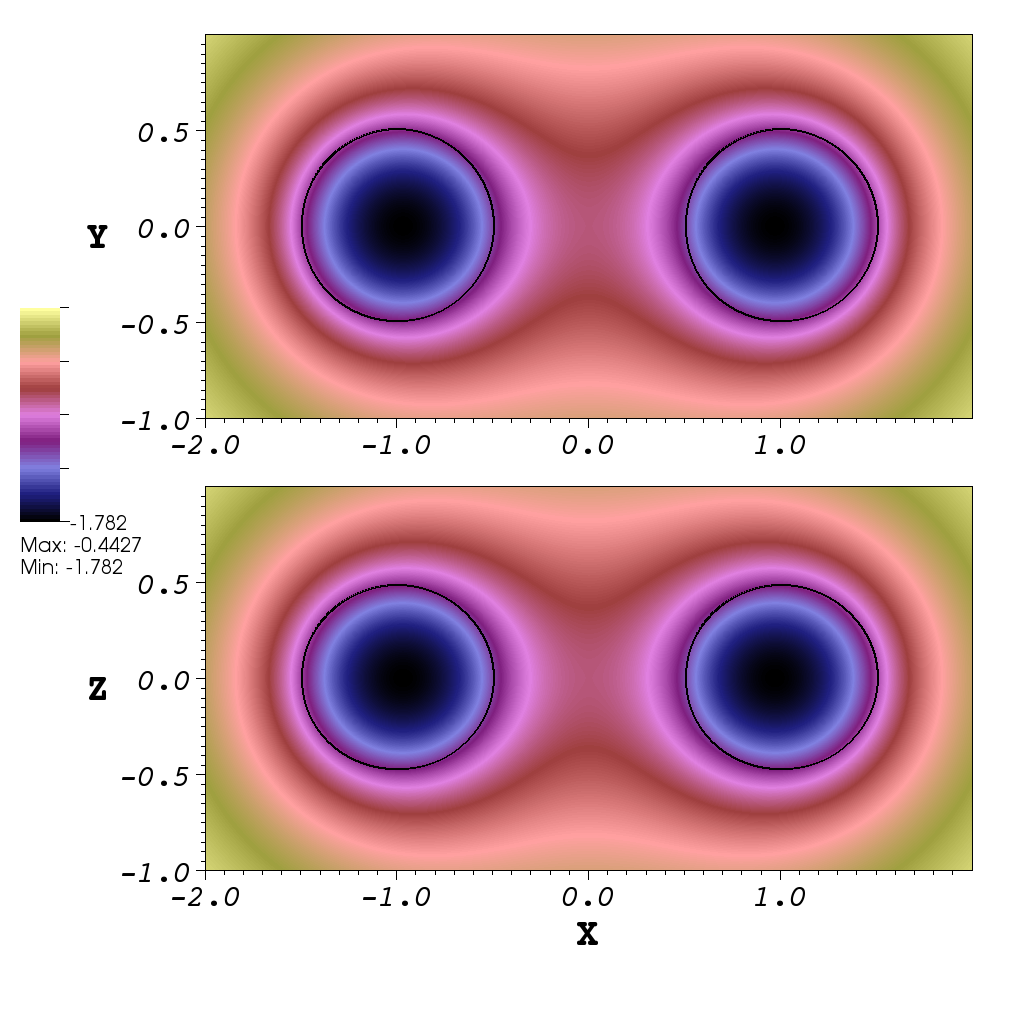}
  \caption{The potential of a homogeneous spheroid binary. 
  Each spheroid has mass density $\rho=1$, eccentricity $e=0.28$, and semi-major axis $a=0.5$. The solid black lines indicate the spheroids' surfaces. 
  The mesh resolution is $768\times384\times384$ cells.
  \label{fig:binary_spheroid_potential}}
\end{figure}

Figure \ref{fig:binary_spheroid_potential} shows the potential for this test problem, 
which is the sum of the potentials of the individual spheroids. 
Thus the analytic solution for this test problem is obtained by modifying the analytic solution found in Section \ref{subsec:homogeneous_spheroid} to account for the shift of the spheroids' centers from the origin. 
This is done by substituting $x-c$ for $x$ in Eqs. \ref{eq:spheroid_potential_inside},  \ref{eq:spheroid_potential_outside}, and \ref{eq:spheroid_potential_auxiliary}, where $c$ is the $x$ coordinate of the center of spheroid.

\begin{figure}[!ht]
  \centering
  \includegraphics[scale=0.4]{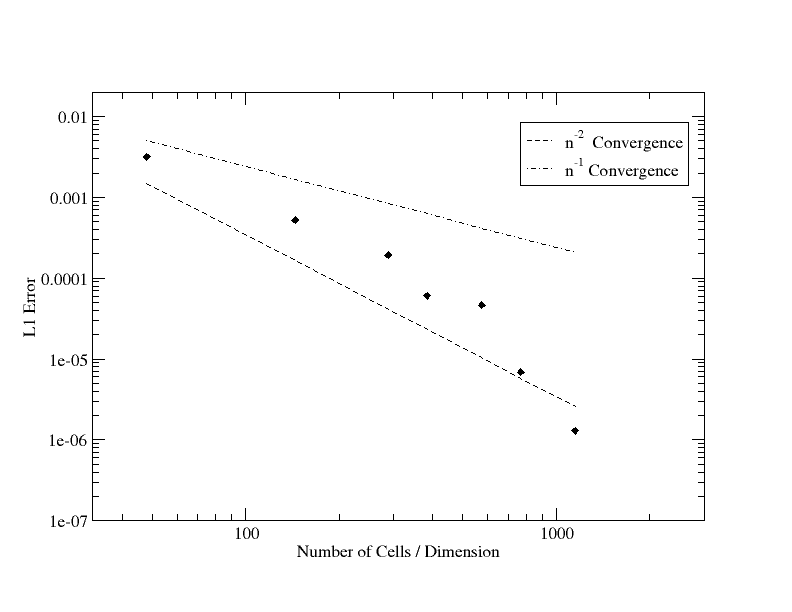}
  \caption{Convergence test of potential for a binary spheroid with uniform mas with the same effective mesh resolutions as the previous test problems.}
  \label{fig:binary_spheroid_convergence}
\end{figure}

As before, we vary the mesh resolution for this test problem to do a convergence test of our solver. 
This is shown in Fig. \ref{fig:binary_spheroid_convergence}. 
The convergence trend is similar to those of the previous test problems, namely, our solver converges better than first order, but does not reach second order convergence. 

\subsection{Scaling}
\begin{figure}[!ht]
  \centering
  \includegraphics[scale=0.4]{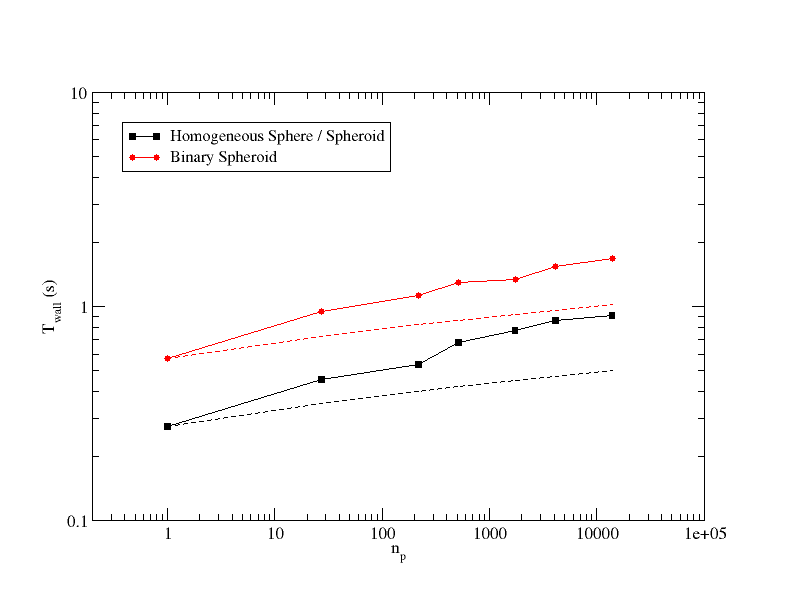}
  \caption{Wall time per solve $T_\mathrm{wall}$ plotted against number of processes $n_p$, demonstrating the weak scaling behavior of \texttt{PSPFFT}. The homogeneous sphere and spheroid test problems assign $48\times48\times48$ cells per process (black solid curve connected with squares), while the binary spheroid test problem assigns $96\times48\times48$ cells per process due to the doubling of computational domain in one dimension (red solid curve connected with circles). The theoretically expected trend for the FFT alone---without communication costs---is shown by the dashed lines, whose vertical offsets are set to match the measured values for $n_p = 1$.}
  \label{fig:weak_scaling}
\end{figure}

We test the weak scaling behavior of our solver by increasing the number of parallel processes while increasing the mesh resolution to maintain a constant amount of work per process. The total CPU time per solve can be expressed 
\begin{equation}
T_\mathrm{CPU} = n_p f(n_p),
\end{equation}
where $n_p$ is the number of processes and $f(n_p) = \mathrm{constant}$ for perfect weak scaling. In Fig. \ref{fig:weak_scaling} we plot the wall time per solve
\begin{equation}
T_\mathrm{wall}  = T_\mathrm{CPU} / n_p = f(n_p)  \label{eq:twall}
\end{equation}
 for the homogeneous spheroid and binary spheroid problems, averaged over 2500 solves (solid lines). The scaling tests were carried out on a Cray XT-4 with quad-core 2.1 GHz AMD Opteron 1354 (Budapest) processors and 8 GB of DDR2-800 memory per node. For compiling the code, we used the vendor-provided PGI compiler version 10 and FFTW version 3.2. Also shown are idealized theoretical trends in the absence of communication costs (dashed lines). For the FFT alone we expect
 \begin{equation}
 T_\mathrm{CPU} = a\, n_c \log n_c,
\end{equation}
where $a$ is a constant and $n_c$ is the total number of cells. Rewriting in terms of $n_c = n_{c/p} n_p$, where $n_{c/p}$ is the number of cells per process, we have
 \begin{equation}
 T_\mathrm{CPU} = b \, n_p \log( n_{c/p} n_p ),
\end{equation}
where $b$ is a new constant. Thus the theoretical expectation we plot is
\begin{equation}
 T_\mathrm{wall} = b \log( n_{c/p} n_p ), \label{eq:twall2}
\end{equation}
with the normalization $b$ set by equating Eq. \ref{eq:twall2} to Eq. \ref{eq:twall} for $n_p = 1$. The number of cells per process is $n_{c/p} = 48^3$ for the homogeneous spheroid and $n_{c/p} = 96 \times 48^2$ for the binary spheroid. We attribute the difference between the measured and theoretically ideal trends to communication costs that rise with the number of processes. We do not consider this extra cost severe, as the time per solve is still about 1 second with $\sim 13,000$ processes.

\section{Conclusion} \label{sec:Conclusion}
We have described our implementation of a parallel solver for Poisson's equation for an isolated system. Our solution method uses Fourier transforms on a distributed unigrid mesh; in particular we use the FFTW library.  
We employ a common protocol, Message Passing Interface (MPI), for communication between processes during a global solve on a distributed-memory system. 
We have shown test problems, numerical convergence, and the weak scaling behavior of our solver. 
We distribute the solver as a library, \texttt{PSPFFT}, which is suitable for use as part of a parallel simulation system. 

\section{Acknowledgements}
This research used resources of the Oak Ridge Leadership Facility at the Oak Ridge National Laboratory, which is supported by the Office of Science of the U.S. Department of Energy (DOE). 
C. Y. C. acknowledges support from the Office of Nuclear Physics and the Office of Advanced Scientific Computing Research of DOE. Oak Ridge National Laboratory is managed by UT-Battelle, LLC, for the DOE. R. D. B. acknowledges support from NSF-OCI-0749204.

\bibliographystyle{unsrt}

\end{document}